\documentclass[runningheads]{svmult}

\usepackage{makeidx}   
\usepackage{graphicx}  
\usepackage{subeqnar}  
\usepackage{multicol}  
\usepackage{physprbb}  
\makeindex             

\newcommand{\greeksym}[1]{{\usefont{U}{psy}{m}{n}#1}}

\newcommand{\uDelta}{\mbox{\greeksym{D}}}

\begin{document}
\title*{Interpreting the Hydrogen IR Lines \protect\newline 
-- Impact of Improved Electron Collision Data}
\toctitle{Interpreting the Hydrogen IR Lines 
\protect\newline -- Impact of Improved Electron Collision Data}
%
%
\titlerunning{Interpreting the Hydrogen IR Lines}
%
\author{Norbert Przybilla\inst{1,2}
\and Keith Butler\inst{3}}
\authorrunning{N.~Przybilla \& K.~Butler}
%
%
\institute{Institute for Astronomy, 2680 Woodlawn Drive, Honolulu, HI 96822, USA
\and Dr. Remeis-Sternwarte Bamberg, Sternwartstr. 7, D-96049 Bamberg, Germany
\and Universit\"atssternwarte M\"unchen, Scheinerstr. 1, D-81679 M\"unchen, Germany}

\maketitle              

\begin{abstract}
We evaluate the effect of variations in the electron-impact excitation cross
sections on the non-LTE line formation for hydrogen in early-type stars. 
While the Balmer lines are basically unaffected by the choice of atomic
data, the Brackett and Pfund series members allow us to
discriminate between the different models. Non-LTE
calculations based on the widely-used approximations of Mihalas, Heasley \&
Auer and of Johnson fail to simultaneously
reproduce the observed optical and IR spectra over the entire parameter range.
Instead, we recommend a reference model using data from {\em ab-initio} calculations up 
to principal quantum number $n$\,$\leq$\,7 for quantitative work. This model is of
general interest due to the ubiquity of the hydrogen spectrum.
\end{abstract}

\section{Introduction}
The quantitative interpretation of the hydrogen line spectrum is one of the
foundations of modern astrophysics. Being the most abundant and most
basic element in the universe hydrogen imprints its signature on the
spectra of stars, nebulae and accretion phenomena. For decades the focus laid
in the modelling of the first members of the Balmer series for
deriving the physical properties of astronomical objects.
In the meantime developments in instrumentation have opened the IR
window to routine observation, which will even gain in importance in the
future, with {\sc Crires} and {\sc Visir} on the {\sc Vlt} being important
cornerstone projects. The next generation of ground-based large telescopes
and the next large space telescope will focus on this wavelength
range, primarily to study the high-$z$ universe. But it will also allow to
investigate local objects in otherwise inaccessible environments, 
e.g. ultra-compact H\,{\sc ii} regions, the Galactic centre and dust-enshrouded 
nearby starburst galaxies. The Brackett and Pfund lines are among the key
diagnostics at these wavelengths. Here we report on the findings of our
reinvestigation of the non-LTE line-formation problem for
hydrogen~\cite{PrBu04}. We confront H\,{\sc i} model atoms of different
degree of sophistication with observations of early-type stars in order to derive a 
set of reference data which is, of course, of much broader interest than for
stellar analyses alone. 

\section{Model calculations}
\begin{figure}[t]
\begin{center}
\includegraphics[angle=-90,width=.495\textwidth]{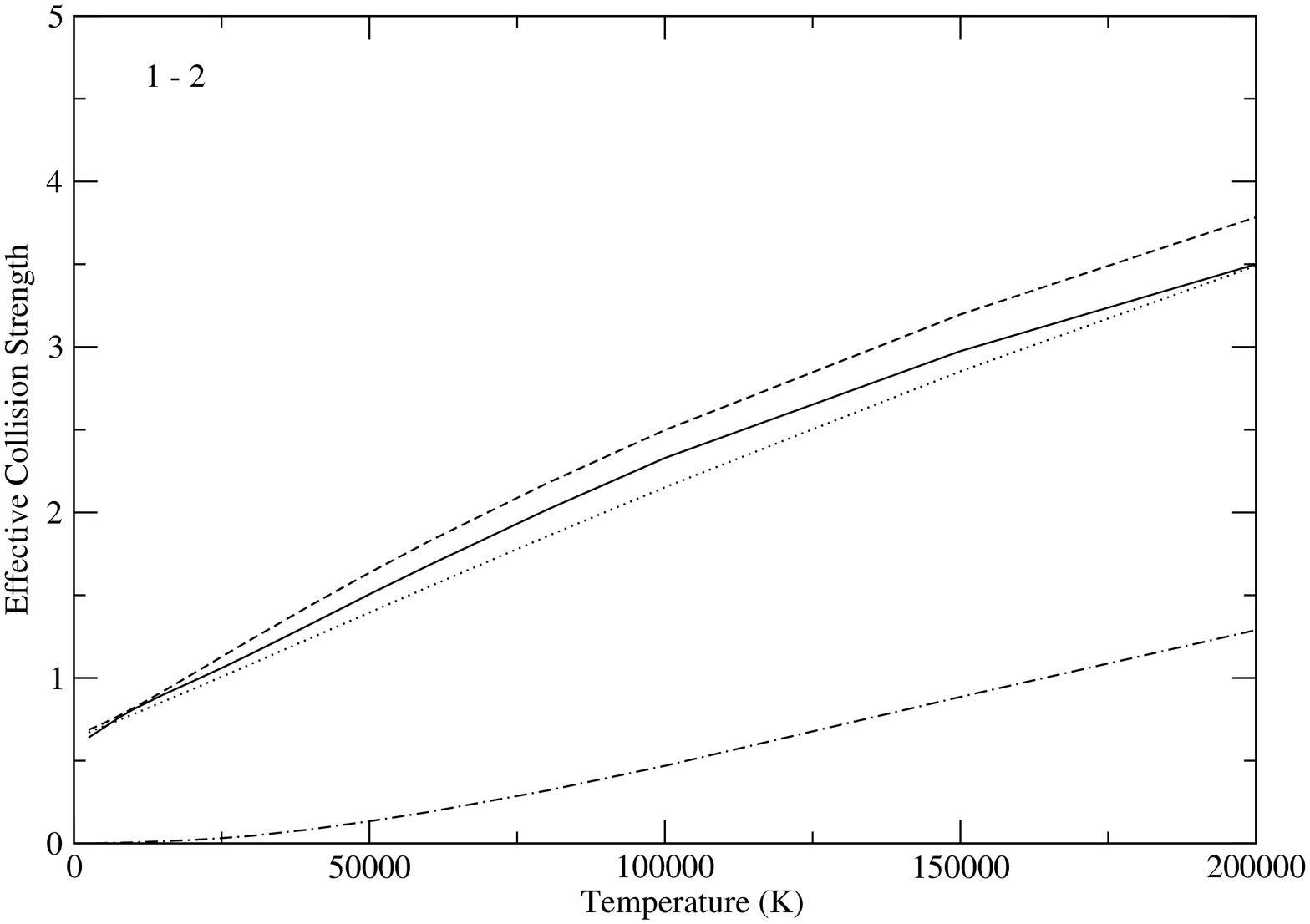}\hfill
\includegraphics[angle=-90,width=.495\textwidth]{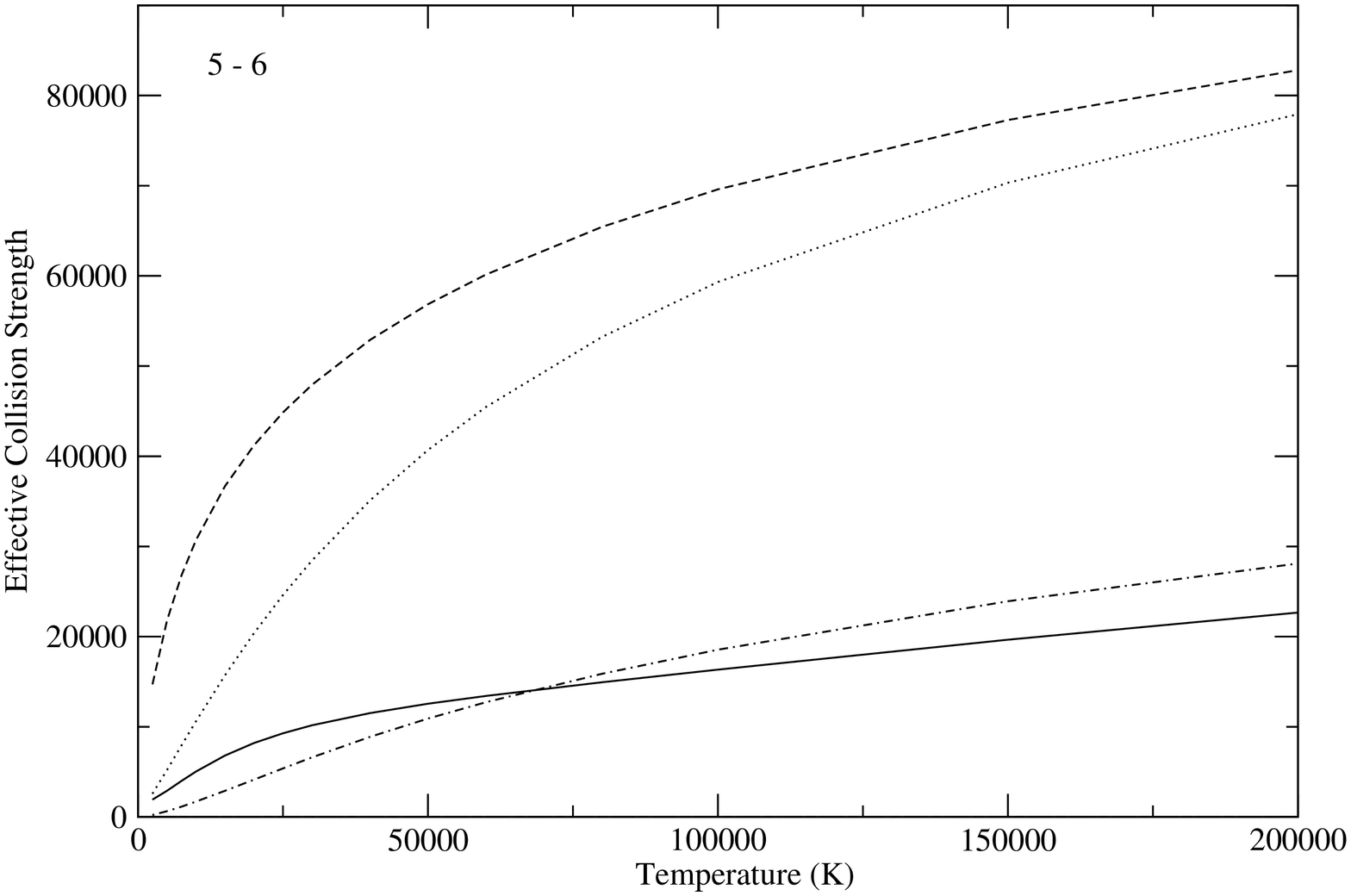}
\end{center}
\caption[]{Comparison of effective collision strengths for transitions
$n$\,--\,$n'$, as indicated. The curves are: B04 (\emph{solid}), J72 (\emph{dotted}), 
MHA (\emph{dashed}), PR (\em{dash-dotted})}
\label{przybillaf1}
\end{figure}

The line-formation computations are carried out using two methods. For
main sequence stars of spectral types later than O and BA-type supergiants a
hybrid approach is chosen. Based on hydrostatic, plane-parallel,
line-blanketed LTE models calculated with the {\sc Atlas9}
code~\cite{Kurucz93}
the non-LTE computations are performed using updated versions of {\sc Detail} and 
{\sc Surface}~\cite{Giddings81,BuGi85}. 
For the modelling of early B and O-type stars we use the non-LTE
model-atmosphere/ line-formation code {\sc Fastwind}~\cite{SaReetal97}
which accounts for spherical extension and hydrodynamic
mass-outflow. It has been recently updated to include an approximate
treatment of non-LTE line-blocking/blanketing~\cite{Pulsetal03}.

Several model atoms for H\,{\sc i} have been implemented.
The two-body nature of the hydrogen atom allows the radiative data to be
obtained analytically. On the other hand,
excitation (and ionization) processes involving a colliding particle
require a numerical solution of the resulting three-body Coulomb problem.
Non-LTE computations had to rely on approximation formulae
\cite{MHA75,Johnson72} (MHA, J72) for
determining collisional excitation rates until recently, as only few measured
cross-sections are available. Using data from extensive {\em ab-initio} calculations up
to $n$\,$\leq$\,7~\cite{Butler04} (B04) we realise a third model atom, further
improved by application of the approximation formula of~\cite{PeRi78} (PR) for
determining electron-collision excitation rates for the remaining
transitions with $n$,$n'$\,$\ge$\,5. A comparison of electron collision
strengths in Fig.~\ref{przybillaf1} shows good agreement of the MHA and
J72 approximations with the results of the detailed computation for the 1--2
transition. However, for transitions among levels with higher 
$n$, like 5--6, large differences can occur, which ultimately will manifest themselves
in model spectra via their impact on the rate equations and thus the level populations. 
Line broadening
is accounted for by using the tables of~\cite{StHu99}. In the case of the
Brackett and Pfund lines we apply the theory of~\cite{Griem60}.

\begin{figure}[t]
\begin{center}
\includegraphics[width=.6\textwidth]{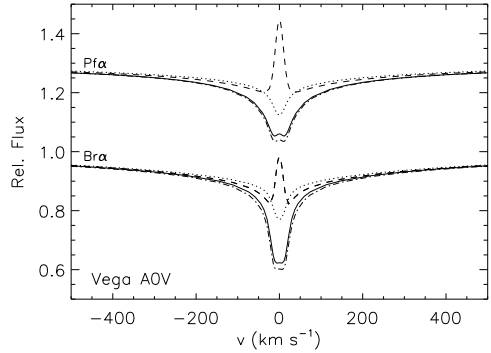}
\end{center}
\caption[]{Comparison of model profiles for Br$\alpha$ and Pf$\alpha$ in
Vega: non-LTE computations using our recommended model atom (electron
collision data of B04$+$PR$+$MHA, \emph{solid}), models using MHA (\emph{dashed})
and J72 data (\emph{dash-dotted}) and LTE profiles (\emph{dotted})
}
\label{przybillaf2}
\end{figure}

The mechanisms driving departures of H\,{\sc i} from detailed balance in
stellar atmospheres have been well understood since the seminal work 
of~\cite{AuMi69a,AuMi69b} (for early-type stars), and numerous
subsequent contributions -- for line formation in the IR e.g. by ~\cite{Zaaletal99}.
The issue here is to study the impact of the {\em local} processes that affect the
radiatively induced departures from LTE, namely collisional interactions, which are 
assumed to be of secondary importance. Indeed, the actual choice of such data
produces no significant differences in the stellar continuum or the Balmer line profiles.
However, consider Fig.~\ref{przybillaf2} where the results from our model
calculations for prominent IR lines in Vega are shown. Apparently, the
choice of collisional data is not a second-order
effect, but a dominant factor for line-formation computations in the IR.
In the following we want to discuss the theoretical background of this behaviour.

\begin{figure}[t]
\begin{center}
\includegraphics[width=.495\textwidth]{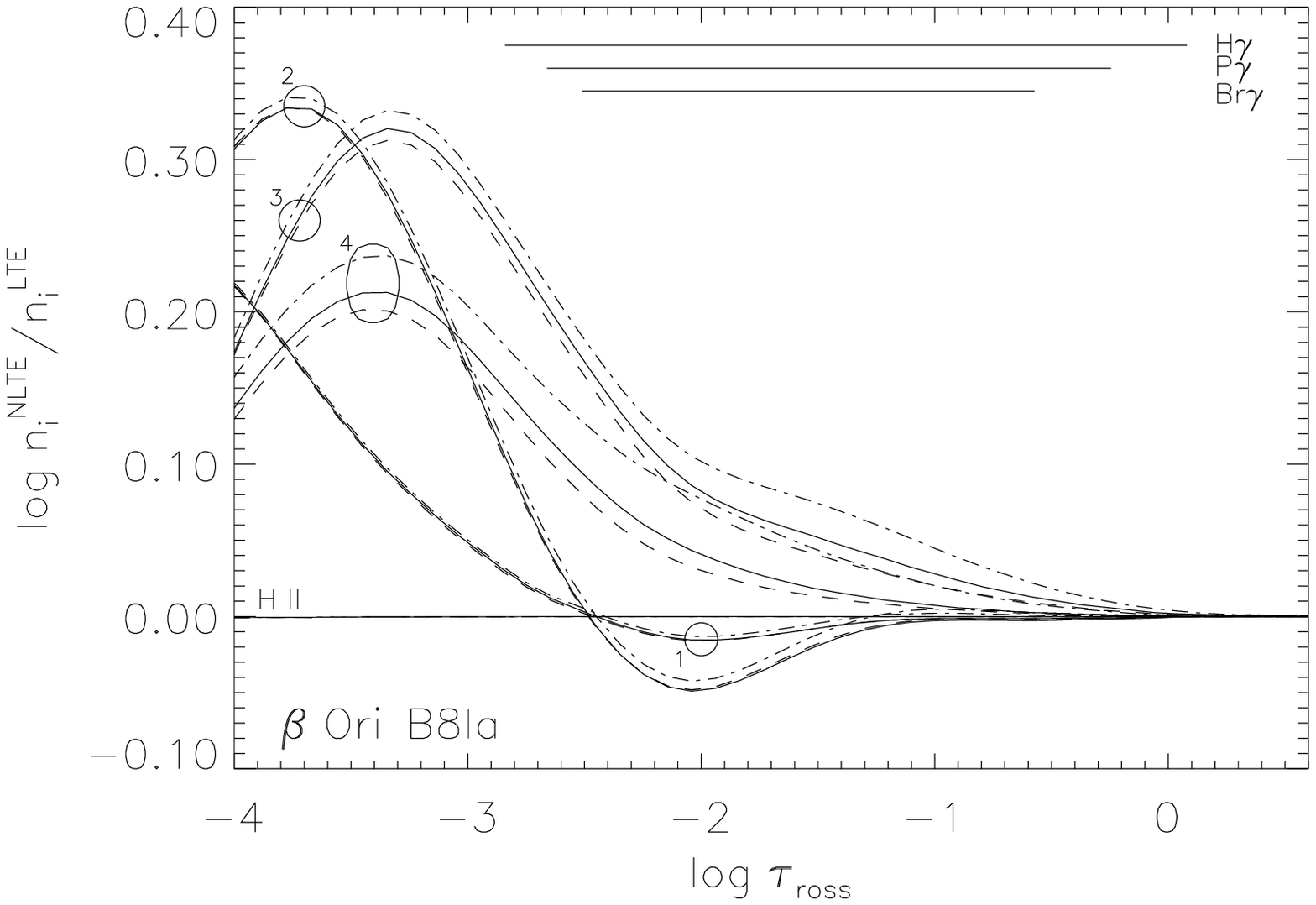}\hfill
\includegraphics[width=.495\textwidth]{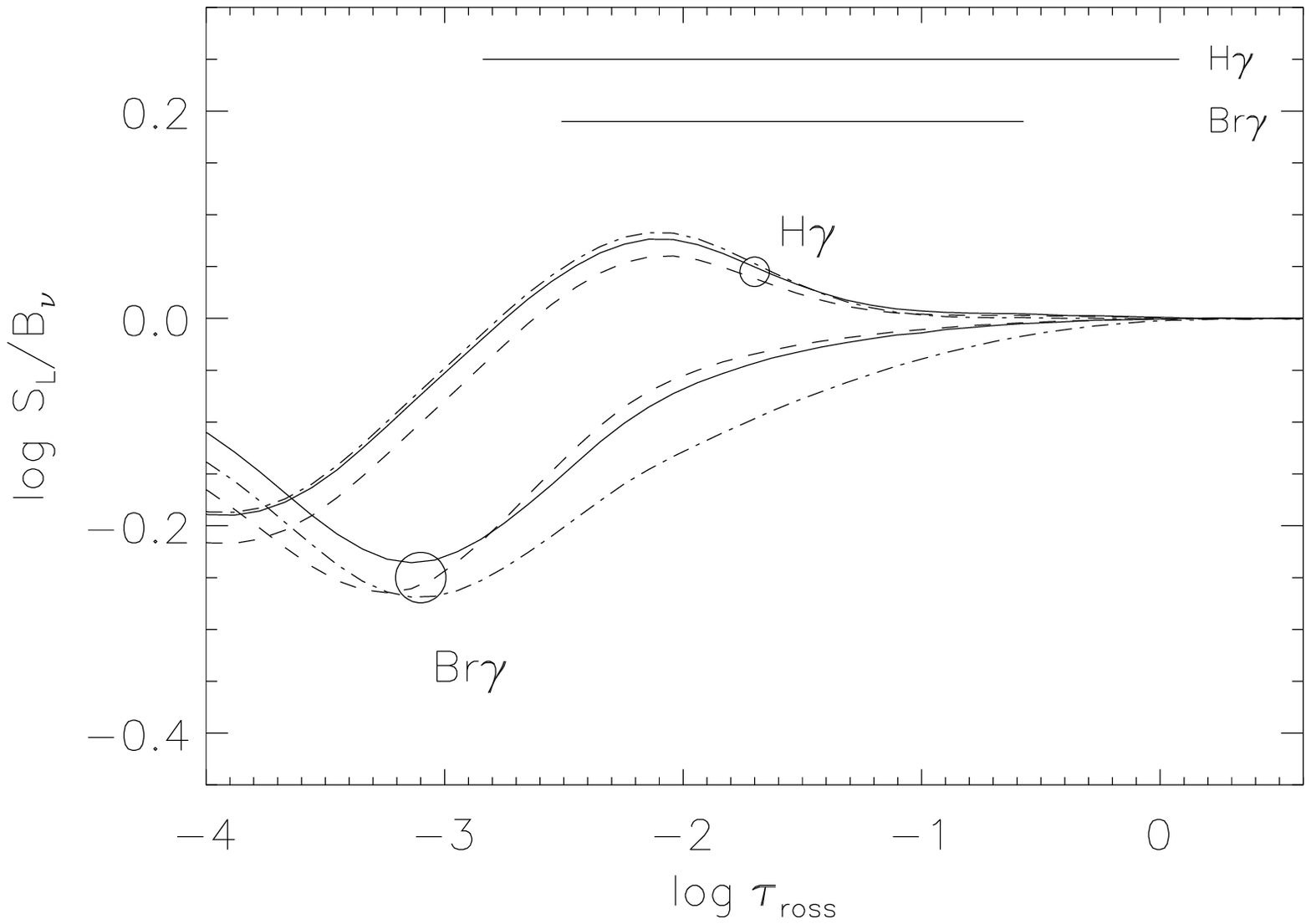}
\end{center}
\caption[]{Run of departure coefficients $b_i$ (left) and ratio of line source
function $S_{\rm L}$ to Planck function $B_{\nu}$ (right) in $\beta$\,Ori as a function
of Rosseland optical depth $\tau_{\rm ross}$. The curves are encoded as in
Fig.~\ref{przybillaf2}. Individual sets of graphs are labelled
by the level's principal quantum number.
Line formation depths for several transitions are indicated}
\label{przybillaf3}
\end{figure}

Departure coefficients $b_i$ for selected levels from computations using 
the different model atoms are
displayed exemplarily for $\beta$\,Ori in Fig.~\ref{przybillaf3}. 
The overall behaviour, i.e. the over- and underpopulation of the levels of 
H\,{\sc i} and of H\,{\sc ii}, is governed by the radiative
processes, while the differences in the collisional data lead to modulations.
These are small for the ground state and become only slightly more pronounced
for the $n$\,$=$\,2 level, as these are separated by comparatively large
energy gaps from the remainder of the term structure. Only colliding particles
in the high-velocity tail of the
Maxwellian velocity-distribution are able to overcome these energy
differences at the temperatures encountered in the star's atmosphere.
However, line-formation computations in the IR
will be affected, as maximum effects from variations of the collisional data
are found for the levels with intermediate $n$ at line-formation depth.
The line source function $S_{\rm L}$ is particularly sensitive to variations of the
ratio of departure coefficients 
\begin{equation}
\left|\uDelta S_{\rm L}\right|=
\left| \frac{S_{\rm L}}{b_i/b_j -\exp(-h\nu/kT)}\uDelta (b_i/b_j)\right|
\approx \left| \frac{S_{\rm L}}{(b_i/b_j -1)+h\nu/kT}\uDelta (b_i/b_j)\right|
\label{przybillae1}
\end{equation}
when $h\nu/kT$ is small. This makes these lines very susceptible to small
changes in the atomic data and details of the calculation.

\section{Confrontation with observation}
\begin{table}[t]
\caption{Stellar parameters}
\begin{center}
\renewcommand{\arraystretch}{1.05}
\setlength\tabcolsep{4pt}
\begin{tabular}{lrrrrrrrrrr}      
\hline
Object & $T_{\rm eff}$ & $\log g$ & $y$ & $R$/R$_{\odot}$ & $v_{\rm turb}$ &
$\dot{M}$ & $v_{\infty}$ & $\beta$\\
 & (K) & (cgs) & & & (km\,s$^{-1}$) & 
(M$_{\odot}$\,yr$^{-1}$) & (km\,s$^{-1}$) & \\
\hline
Vega         &  9550 & 3.95 & 0.09  & 2.8  & 2 & {\ldots} & {\ldots} & {\ldots}\\
$\beta$\,Ori & 12000 & 1.75 & 0.135 & 104  & 7 & {\ldots} & {\ldots} & {\ldots}\\
$\tau$\,Sco  & 31400 & 4.24 & 0.09  & 5.1  & 3 & 9.00$\cdot$10$^{-9}$ & 2000 & 2.4/2.5\\
HD\,93250    & 46000 & 3.95 & 0.09  & 15.9 & 0 & 3.45$\cdot$10$^{-6}$ & 3250 & 0.9\\
\hline
\end{tabular}
\end{center}
\label{przybillat1}
\end{table}

\begin{figure}[ht!]
\begin{center}
\includegraphics[width=.484\textwidth]{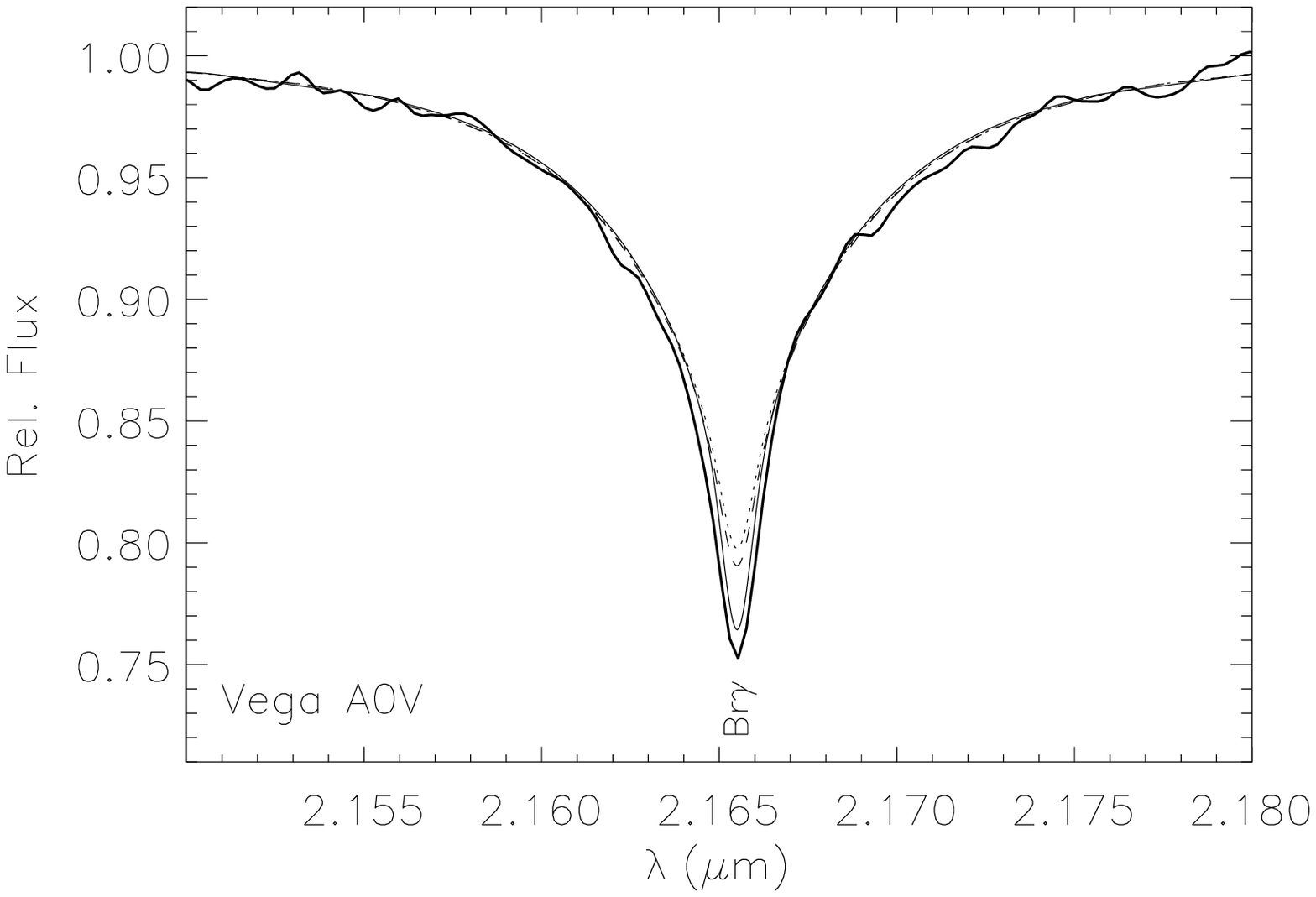}\hfill
\includegraphics[width=.484\textwidth]{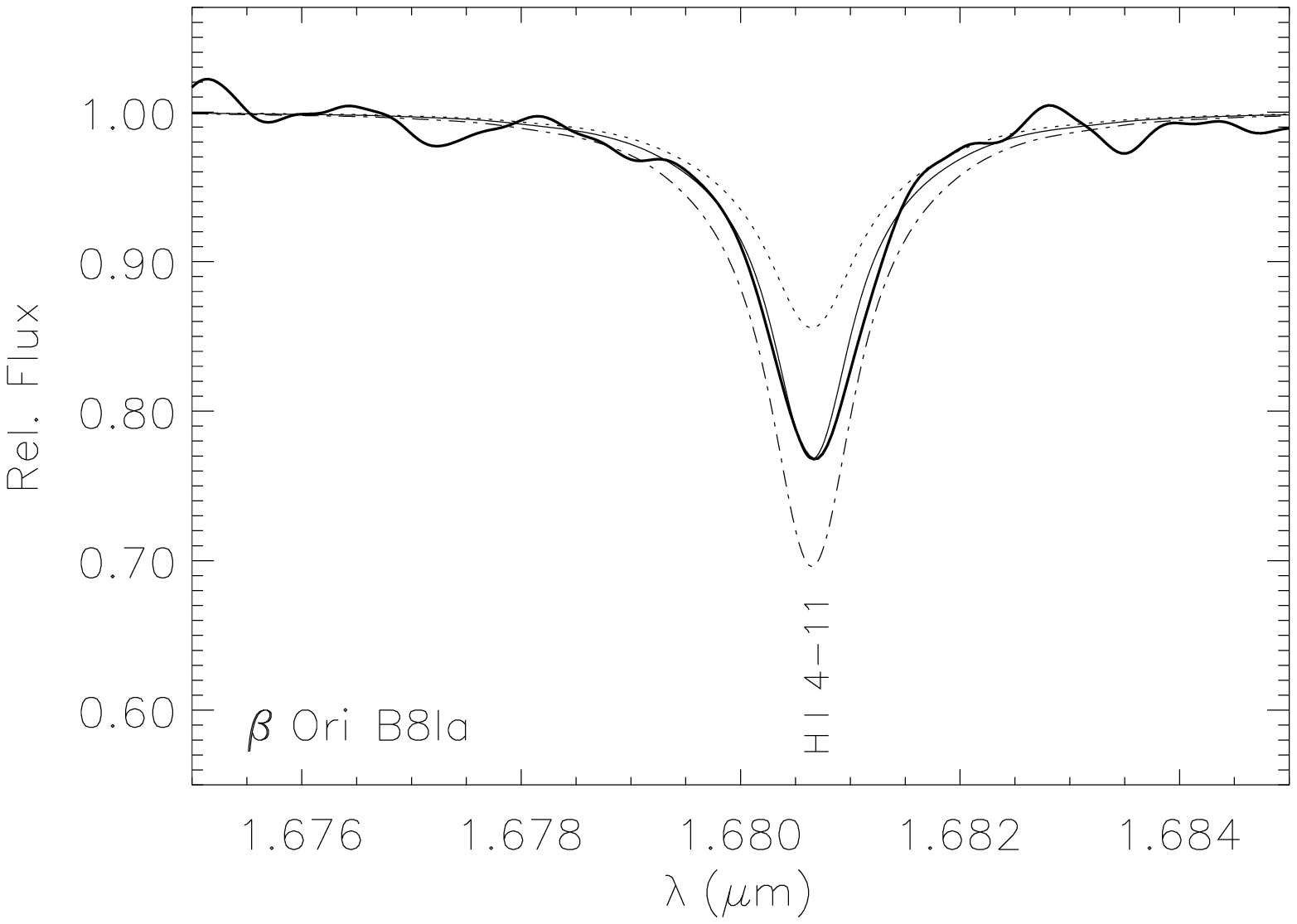}
\end{center}
\caption[]{Comparison of selected synthetic spectra (encoded as in
Fig.~\ref{przybillaf2}) with observed IR hydrogen lines (\emph{thick
solid}). LTE modelling fails to reproduce the observations.
Model atoms relying on the MHA approximation underestimate the
NLTE strengthening of the Br$\gamma$ line core in the main sequence star (left). Those
using the J72 approximation predict too strong features in BA-type supergiants
(right). Our recommended model provides good fits consistently throughout the high
and low density regimes \vspace{-4.5mm}}
\label{przybillaf4}
\begin{center}
\includegraphics[width=.484\textwidth]{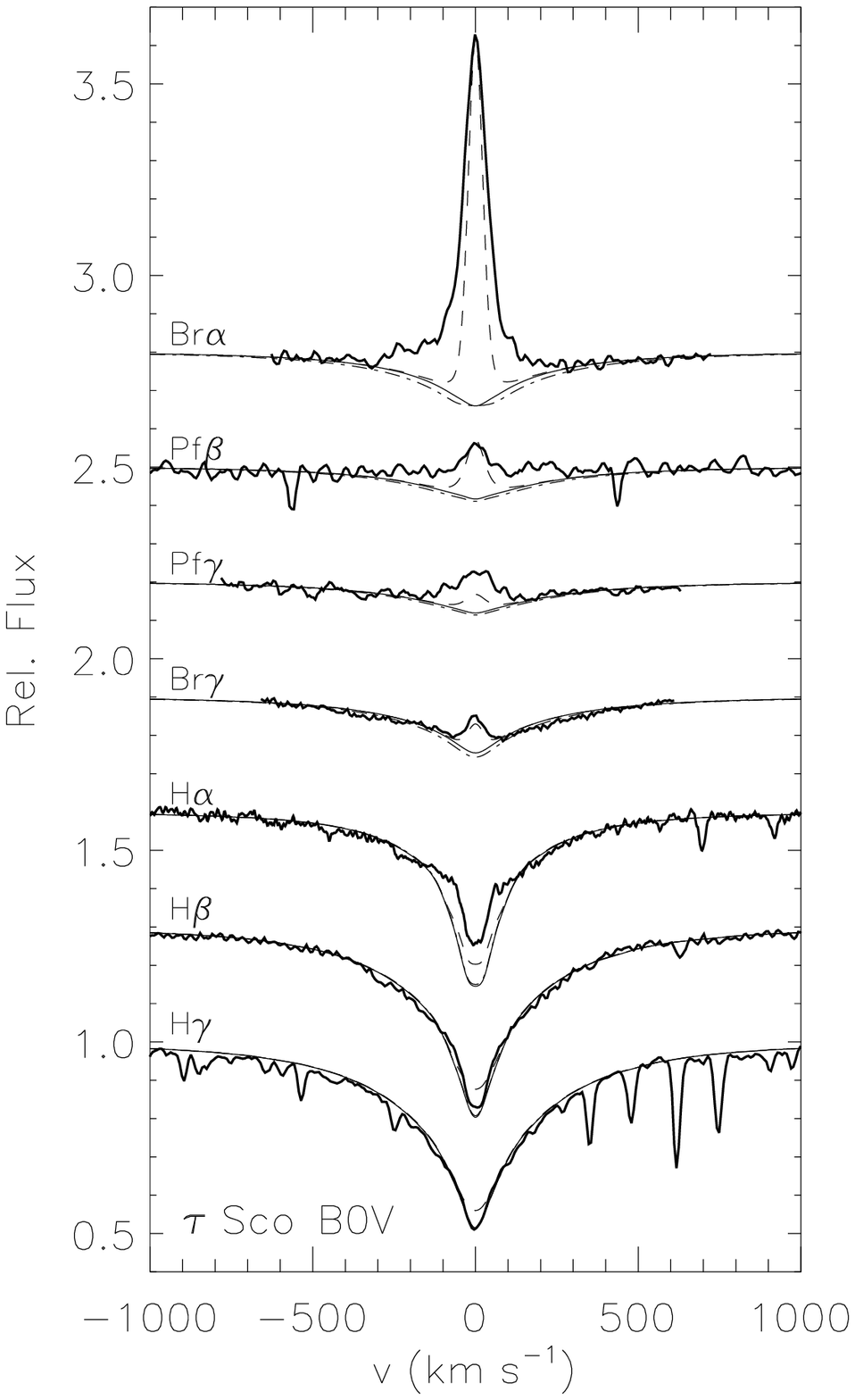}\hfill
\includegraphics[width=.484\textwidth]{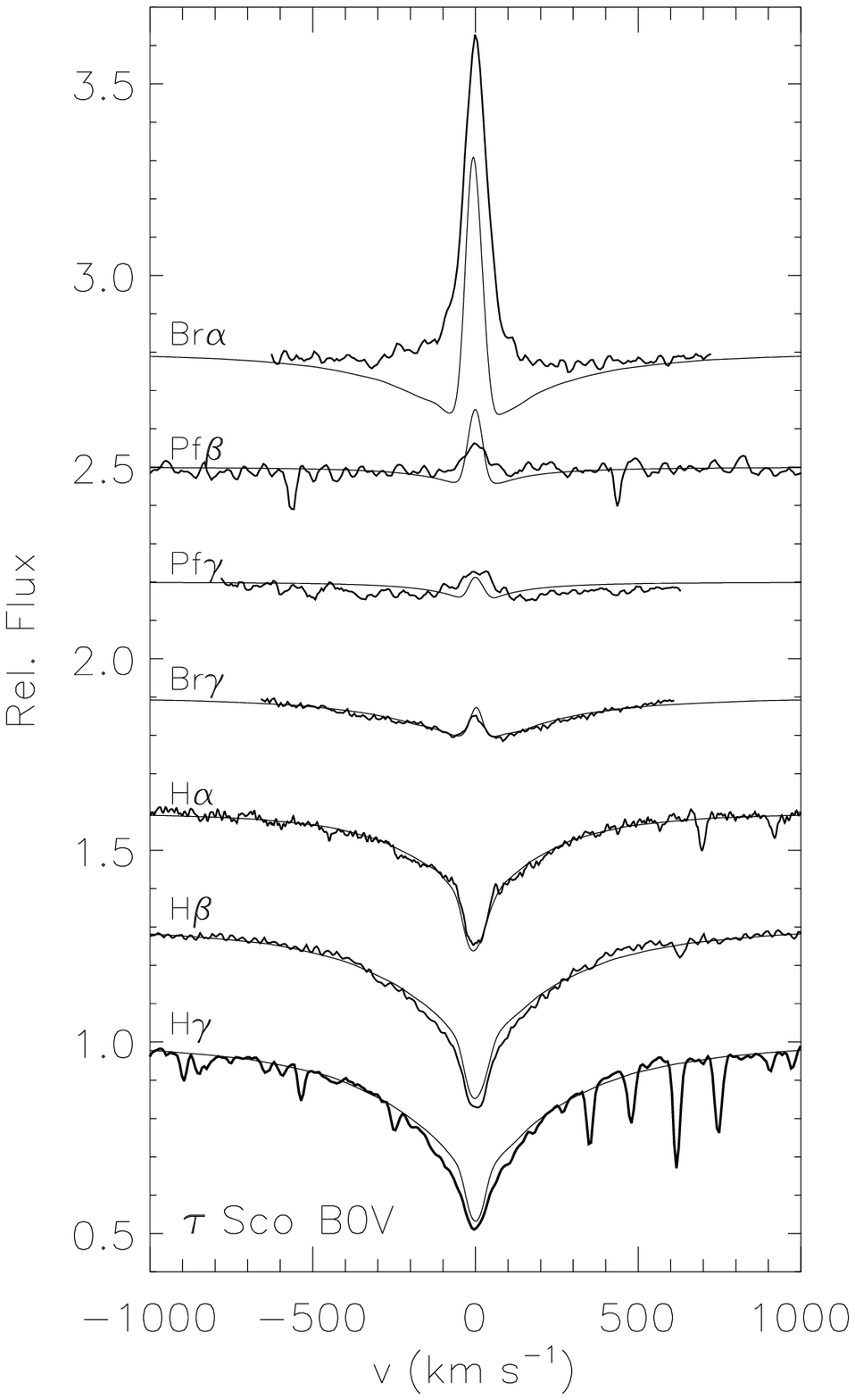}
\end{center}
\caption[]{Spectrum synthesis for hydrogen lines in $\tau$\,Sco (encoded as
in Fig.~\ref{przybillaf2}). Left: hydrostatic computations; using the MHA
approximation we reproduce the findings of \cite{Zaaletal99}, i.e.
obtain a fair fit of the IR core-emission peaks. However, the other model
implementations indicate no emission cores. Right: hydrodynamic approach;
slight differences in the atmospheric structure at line-formation depths
suffice to improve the quality of the line fits. Notable residuals are
still present for Br$\alpha$, a consequence of the sensitivity
of the calculation to the non-LTE amplification close to population inversion} 
\label{przybillaf5}
\end{figure}

\begin{figure}[t]
\begin{center}
\includegraphics[width=.6\textwidth]{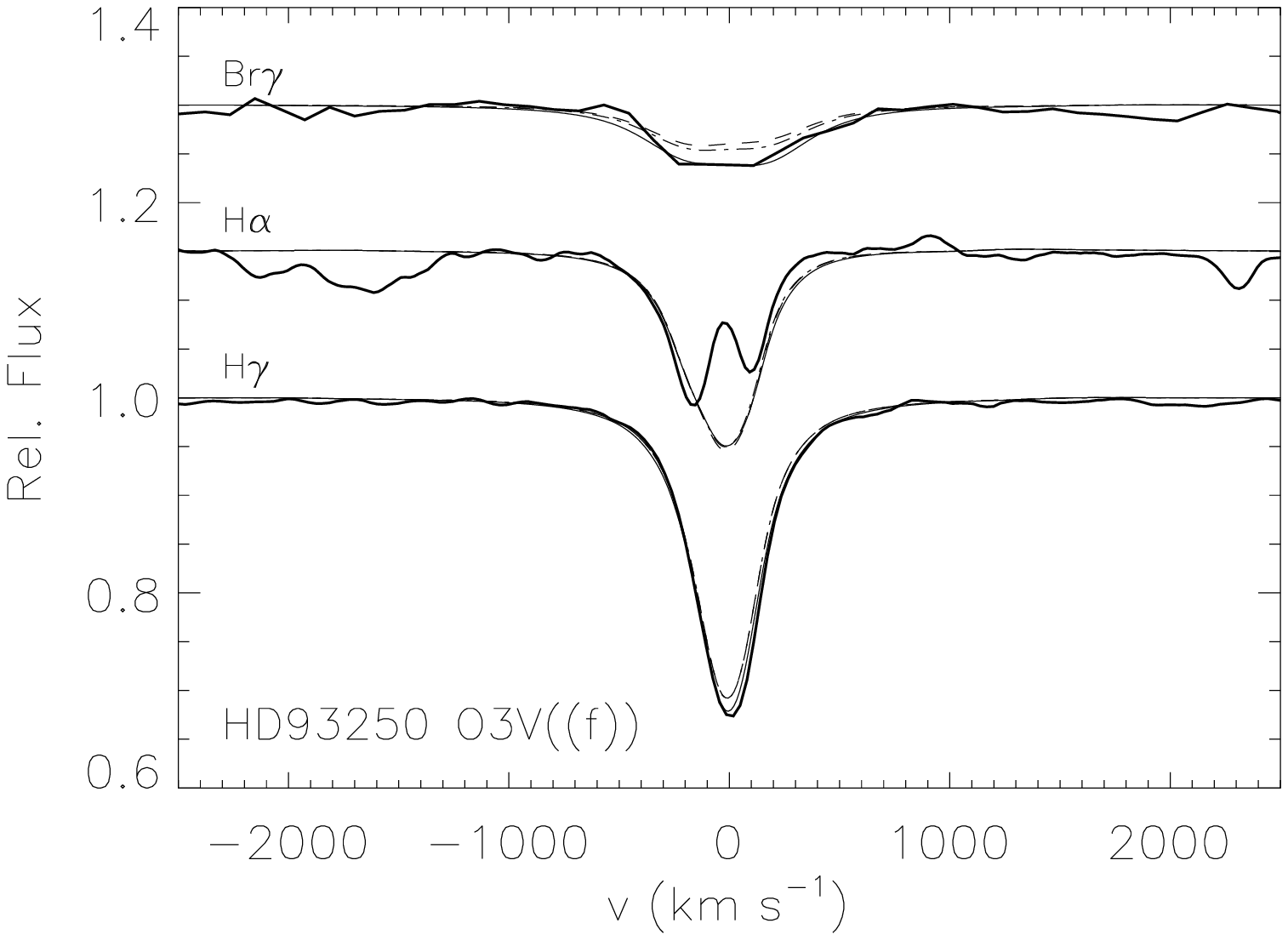}
\end{center}
\caption[]{Spectrum synthesis for an early O-type star (encoded as in
Fig.~\ref{przybillaf2}). Accurate electron collision data are mandatory for
consistent modelling of the Brackett and Balmer lines. 
H$\alpha$ shows some residual nebular emission}
\label{przybillaf6}
\end{figure}

In order to determine which model atom is suited best for quantitative
studies we compare model predictions for IR lines with observations for a few objects
sampling the parameter space, see Figs.~\ref{przybillaf4}--\ref{przybillaf6}. 
This has to rely on a variety of sources for 
the observations, see~\cite{PrBu04} for details. The stellar parameters
adopted for the model calculations are summarised in Table~\ref{przybillat1}
($y$ is He abundance by number and $\beta$ the wind velocity parameter).

We conclude from this comparison that model atoms using the MHA
approximation overestimate the thermalising effects of electron collisions,
thus dampening the non-LTE effects (as determined in the high density
environment of the main sequence star Vega). On the other hand, use of the
J72 approximation leads to an unrealistic non-LTE strengthening in
supergiants, like in the case of $\beta$\,Ori. In both cases good agreement
between theory and observation is obtained using the superior atomic data from
\emph{ab-initio} computations (Fig.~\ref{przybillaf4}). The IR lines of the
early B-type star $\tau$\,Sco turn out to react highly sensitive to
the atmospheric conditions (Fig.~\ref{przybillaf5}), as they are subject to
considerable non-LTE amplification (Eqn.~\ref{przybillae1}). Finally,
HD\,93250 may act as a benchmark for the study of objects at the earliest
phases of stellar evolution of massive stars. Here, the differences in the
predicted equivalent widths from the different model atoms may reach a factor 
$\sim$2 (Fig.~\ref{przybillaf6}). The Balmer lines remain basically unaffected 
by the choice of the atomic data in all the cases studied. 

\clearpage

\section{Recommendations}
Use of electron collision data from \emph{ab-initio}
computations is \emph{mandatory} in order to 
derive consistent results from the diagnostic lines in the visual and IR.
We recommend the data of~\cite{Butler04} for the evaluation of collision rates of
H\,{\sc i}, supplemented by the approximation formulae by~\cite{PeRi78} and
of~\cite{MHA75} for those
transitions not covered by the {\em ab-initio} computations. This applies
not only to the modelling of early-type stars as we have done but to all
hydrogen plasmas.

%

\end{document}